# Spectroscopic properties and lattice dynamics of ferroelectric and related functional oxide ceramics

Jan Petzelt[a] and Stanislav Kamba[a]

Dielectric response as a function of frequency for high-permittivity dielectric and ferroelectric materials is discussed emphasizing the dynamic behaviour in the microwave and infrared range. After introducing the usual modelling of the polar phonon response and anharmonic hopping in locally dynamically disordered solids, including discussion of displacive and order-disorder ferroelectric phase transitions, we summarised the experimental data for selected ferroelectric, incipient ferroelectric and antiferroelectric perovskite ceramics (STO, BTO, BST, PZO), discussing the dielectric grain size effect due to a low-permittivity (dead) grain-boundary layer. Other important structural types of ferroelectrics (Aurivillius compounds SBT, pyrochlores CNO, PMN) are mentioned, as well. Attention is then paid to relaxor ferroelectrics (PLZT, PMN, PMT, PST, NBT of perovskite structure and SBN and a novel BLNTN solid solution system and SLTN of tungsten-bronze structure) and their extremely broad and complex dielectric dispersion. Then some new results on magnetoelectric multiferroics ($BiFeO_3$, $BiFe_{1/2}Cr_{1/2}O_3$, $EuTiO_3$) are summarised and finally our results on microwave ceramics (used for microwave applications) are noted, discussing mainly the problem of extrapolating the microwave dielectric properties from the infrared and THz range. A glossary is added, briefly explaining the meaning of several concepts which might not be quite familiar for the reader.

1. Introduction

In this review we shall limit ourselves on discussing the linear dielectric properties of non-conducting oxide ceramics with high dielectric constant from the point of view of lattice dynamics. Linear dielectric response (response of the electric displacement $\boldsymbol{D}$ on the electric field $\boldsymbol{E}$) is described by the complex relative permittivity (dielectric function) tensor $\varepsilon_{ij}^*(\omega) = \varepsilon_{ij}'(\omega) + i\varepsilon_{ij}''(\omega)$ as a function of angular frequency $\omega$. Since we shall neglect it's spatial dispersion (e.g. optical activity), $\varepsilon_{ij}(\omega)$ is a symmetric second–rank tensor. For a brief more general discussion of the dielectric function see e.g. [1]. As for all causal response functions (i.e. when the response cannot precede the action), the real and imaginary parts of the permittivity are related by the integral Kramers-Kronig relations:

$$\varepsilon_{ij}'(\omega) - \delta_{ij} = \frac{2}{\pi} P \int_0^{+\infty} \frac{x \varepsilon_{ij}''(x)}{x^2 - \omega^2} dx$$

$$\varepsilon_{ij}''(\omega) - \frac{\sigma_{ij}(0)}{\varepsilon_0 \omega} = -\frac{2\omega}{\pi} P \int_0^{+\infty} \frac{\varepsilon_{ij}'(x) - \delta_{ij}}{x^2 - \omega^2} dx \quad (1)$$

where the dc conductivity $\sigma_{ij}(0)$ was introduced to account for materials with nonzero dc conductivity, $\varepsilon_0$ is permittivity of the free space and $\delta_{ij}$ is the Kronecker symbol. $\boldsymbol{P}$ means the principal value of the integral. Here we will limit ourselves to the discussion of principal values of the dielectric tensor components, which for orthorhombic and higher symmetry crystals, are obtained for $\boldsymbol{E}$ along the crystallographic axes. For cubic materials the permittivity is isotropic (one principal value of permittivity), for optically uniaxial and biaxial crystals there are two and three principal permittivity values, respectively. In the case of polycrystals with anisotropic crystallites like ceramics or polycrystalline films, the sample is usually macroscopically isotropic (if no texture of the microstructure is considered) and the corresponding single effective dielectric response component should be modelled by some appropriate averaging of the principal crystal values. This is non-trivial since one has to consider the depolarizing fields acting on the individual grains, which depend on the shape and topology of the grains. We shall discuss the appropriate models based on the Effective Medium Approximation (EMA) later and hereafter we shall neglect the subscripts *ij*.

From the point of view of basic understanding of the physical mechanisms of polarization as well as for applications, the main interest is to describe and understand the dispersion regions in the $\varepsilon(\omega)$ spectra where $\varepsilon'(\omega)$ changes with the frequency ω. Usually, they consists of peaks in the dielectric loss function $\varepsilon''(\omega)$ and the corresponding (via Kramers-Kronig relation) dispersion regions in the real permittivity $\varepsilon'(\omega)$. The complete dielectric response may be described by additive contributions of such dispersion regions. Let us start from the highest frequencies (optical VIS and UV range) given by the electron absorption processes. Since the electron transitions involve the whole Brillouin zone, no simple phenomenological models with a simple physical meaning used for the description of the complex dispersion are available. However, the optical electronic contributions are usually not dominant (except for elemental dielectrics and semiconductors like diamond, Si, Ge, which show no first-order phonon absorption processes) for understanding the dielectric properties. Their contribution to the relative permittivity (index of refraction squared) is usually <10. Therefore we shall omit their discussion and take their contribution to permittivity as a (possibly only slightly temperature dependent) constant $\varepsilon_\infty$.

For crystals, in which the primitive unit cell contains at least two atoms, the infrared (IR) spectrum is dominated usually by one-phonon absorption. The IR active phonons have a non-zero dipole moment associated with their vibrations (so called polar modes) and their contribution to the dielectric dispersion can be usually well described by a classical damped harmonic oscillator model. In the case of more complex structures with more vibrational degrees of freedom in the unit cell, the simplest model for the dielectric function is

$$\varepsilon^*(\omega) = \varepsilon_\infty + \sum_j \frac{\Delta\varepsilon_j \omega_j^2}{\omega_j^2 - \omega^2 + i\omega\gamma_j} \quad (2)$$

with $\omega_j$, $\gamma_j$ and $\Delta\varepsilon_j$ the $j$-th transverse optic (TO) mode frequency, damping and dielectric strength (contribution to the static permittivity), respectively. Only TO modes are active in the first–order absorption processes since the electric field in the IR wave is also transverse. The first-order absorption of the IR wave by a phonon wave is possible only if both waves propagate in the same direction and have the same wavelength (same wave vector), so that they can interact over a long distance. This model does not account for mode coupling phenomena, which may become important if the corresponding loss functions from the same spectrum (same symmetry of the excitation) overlap. The corresponding formula for two linearly coupled oscillators is [2]

$$\varepsilon^*(\omega) = \varepsilon_\infty + \frac{\Delta\varepsilon_1\omega_1^2(\omega_2^2 - \omega^2 + i\omega\gamma_2) + \Delta\varepsilon_2\omega_2^2(\omega_1^2 - \omega^2 + i\omega\gamma_1) - 2\sqrt{\Delta\varepsilon_1\omega\Delta\varepsilon_2}\,\omega_1\omega_2\alpha}{(\omega_1^2 - \omega^2 + i\omega\gamma_1)(\omega_2^2 - \omega^2 + i\omega\gamma_2) - \alpha^2}$$

(3)

The coupling constant $\alpha$ can be generally complex. Roughly speaking, its real part renormalizes the mode frequencies and the imaginary part deforms the mode spectral band-shapes.

In the IR range, the most usual technique to measure the dielectric function is normal-incidence specular power reflectivity $R(\omega)$ from mirror-flat opaque samples, which is related to the dielectric function $\varepsilon(\omega)$ by

$$R(\omega) = \left|\frac{\sqrt{\varepsilon(\omega)} - 1}{\sqrt{\varepsilon(\omega)} + 1}\right|^2 \quad (4)$$

Since the power reflectivity yields only one spectrum (the information about the phase of the reflected radiation is missing in the standard FTIR spectroscopy), a model fitting procedure (or Kramers-Kronig analysis) is necessary to determine the complex dielectric function $\varepsilon(\omega)$.

The reflectivity spectra are sensitive not only to features near TO frequencies (maxima in the reflectivity in the case of not very strong and weakly damped polar modes), but also to longitudinal optic (LO) frequencies (minima in the reflectivity in the case of weakly damped modes). Therefore, for the successful fitting of the reflectivity spectra a more general dielectric function in factorized form was introduced [3], so-called generalized multi-oscillator formula

$$\varepsilon^*(\omega) = \varepsilon_\infty \prod_{j=1}^{n} \frac{\omega_{LOj}^2 - \omega^2 + i\omega\gamma_{LOj}}{\omega_{TOj}^2 - \omega^2 + i\omega\gamma_{TOj}} \quad (5)$$

which introduces an independent damping parameter $\gamma_{LOj}$ of the *j*-th LO mode. This formula does not guarantee automatically physically acceptable parameters (i.e. positive dielectric losses for all frequencies), but accounts partially for the mode coupling phenomena, because each oscillator is fitted by 4 free parameters unlike the independent classical oscillator, which is determined by 3 parameters. This formula (eqn 5) yields in the static limit the generalized Lyddane-Sachs-Teller relation,

$$\frac{\varepsilon(0)}{\varepsilon_\infty} = \prod_{j=0}^{n} \frac{\omega_{LOj}^2}{\omega_{TOj}^2} \quad (6)$$

which is frequently used to compare the lattice contribution to the permittivity with the actually measured (by capacitance technique) low-frequency $\varepsilon(0)$.

In an ideal harmonic crystal, the only additional dielectric dispersion below the TO phonon frequencies is due to possible piezoelectric resonances (permitted by symmetry only in non-centrosymmetric structures), which describes the contribution to permittivity accounting for the difference between the smaller mechanically clamped permittivity and higher stress-free permittivity. Usual (i.e. weak) lattice anharmonicities introduce additional multiphonon absorption, which provides THz and microwave (MW) tails to the one-phonon dielectric loss spectra. But these mechanisms are mostly too weak to account for appreciable dispersion in the real permittivity below the IR phonon range. However, many materials, particularly those exhibiting high permittivity (of the order of $10^2$ and more), show additional dispersion of the relaxational type, which is specified by a monotonous decrease in the permittivity with increasing frequency) in the MW and lower-frequency range. This can be of strongly anharmonic lattice origin, if some of the lattice ions occupy two or more not very distant (0.1-1 Å) lattice sites. Because of thermal fluctuations, the ions hop among the permitted sites and their local hopping usually contributes to the dielectric function by a Debye relaxation

$$\varepsilon^*(\omega) = \varepsilon_{ph}(\omega) + \frac{\Delta\varepsilon_R \omega_R}{\omega_R + i\omega} \qquad (7)$$

where $\Delta\varepsilon_R$ is the dielectric strength of the relaxation and $\omega_R$ is the temperature dependent relaxation frequency which corresponds to the peak value in the loss spectra. Qualitatively similar dynamics can be also caused by charged defects and dopants, particularly if they are of smaller ionic size than the substituted ions and differ from their valence (aliovalent substitution).

In partially disordered materials, the total dielectric dispersion below the phonon range is frequently even more complex and requires fitting with several Debye relaxations or with other models (mostly phenomenological) which correspond to some continuous distribution of Debye relaxations. Such modelling is mostly used for glasses, polymers and other soft matter, which is substantially disordered. Therefore, it will not be discussed here. However, we would like to emphasize the existence of a very strong and complex dielectric dispersions existing in dipolar glasses and particularly in so called relaxor ferroelectrics [4]. These are crystalline materials (mostly chemically heterogeneous like solid solutions) with an essentially periodic lattice, but additional nanosized polar regions which are specified by small ionic displacements from their non-polar periodic lattice sites. Their own coherent dynamics depends strongly on temperature and is of a complex relaxational type. At high temperatures, where such regions form, the relaxational dielectric dispersion separates from the normal THz phonon dynamics and softens (its frequency decreases) on cooling through the MW range down to arbitrarily low frequencies, at the same time causing extreme broadening over the whole loss spectra. Typically, at cryogenic temperatures in all these materials the dielectric loss spectra remain nonzero and are frequency independent up to the MW range, creating the so called 1/f or flicker noise - see below.

Let us make a short additional remark on materials in which the conductivity is not negligible. For good conductors, the ac conductivity response can be usually quite well described by a Drude-type dispersion, which is identical to damped harmonic oscillator response with a zero eigen-frequency (zero restoring force, free charge carriers). For such a response the ac conductivity is a decreasing function of frequency and its contribution to static permittivity via Kramers-Kronig relations is negative. However, in poor conductors (e.g. with dominating hopping conductivity) or in systems with inhomogeneous conductivity (which is quite usual in ceramics where the conductivity in the bulk grain and grain boundary differ), the ac conductivity is - at least in some spectral range - an increasing function of frequency. From Kramers-Kronig relations (eqn 1) it follows [5] that the conductivity contribution to the static permittivity in such cases is positive and could be quite pronounced. In such materials one might observe huge dielectric dispersions in the low-frequency range (also often following the Debye relaxation model) giving rise to so called 'giant' permittivity materials [6]. In case of ceramics or core-shell composites, in which the grain boundaries (shells) are more resistive than the grain bulk (core), this effect can be applied in so called barrier-layer capacitors [7]. The dielectric dispersion region in such systems is usually limited to relatively low frequencies (below MHz) and is strongly temperature dependent, vanishing at low temperatures, since the conductivity is, as a rule, thermally activated. We shall not discuss here anymore these complex phenomena, since, from the point of view of lattice dynamics, they are not intrinsic.

Concerning the dynamic dielectric behaviour, a specific attention is required to ferroelectric phase transitions. It has been known for 50 years [8] that displacive ferroelectric phase transitions are characterized by a soft polar phonon mode which in the paraelectric phase (as a rule above the Curie temperature $T_C$), accounts for the whole value of permittivity. It is clear from Eq. (6) that softening of one TO phonon mode to zero provides a divergence of the static permittivity, if the other phonon parameters do not critically depend on temperature (which is expected and actually was always experimentally confirmed). So the classical softening Cochran law [8]

$$\omega_{SM}^2 = A(T - T_C), \qquad (8)$$

yields as a consequence the Curie-Weiss law for the static permittivity

$$\varepsilon_0 = \varepsilon_B + C/(T - T_C) \qquad (9)$$

with the Curie-Weiss constant $C = \Delta\varepsilon_{SM}\omega_{SM}^2/A$. $\varepsilon_B$ is the sum of the electronic contributions and contributions from other (so called hard) TO modes which can be considered to be temperature independent.

The ferroelectric order-disorder transitions, on the other hand, appear in highly anharmonic crystals, where one disordered ion type assumes two or more positions with the same probability in the paraelectric phase and the hopping among them can be usually modelled by a critical Debye relaxation (Eq. 7), whose frequency slows down linearly to zero

$$\omega_R = A_R(T - T_C) \qquad (10)$$

which again in the static limit yields the Curie-Weiss law for the permittivity. It should be noted that, in the displacive case if the soft phonon damping is essential, the soft mode becomes overdamped (its damping $\gamma > 2\omega_{SM}$) close to $T_C$ and it may become very difficult to determine experimentally the natural soft mode frequency $\omega_{SM}$. In fact, the response of a heavily overdamped oscillator approaches that of the Debye relaxation in the frequency range of the dielectric-loss maximum and below it [9]. In this case the maximum in the dielectric loss spectrum is more a representative frequency (easily experimentally determined), which approaches the relaxation frequency $\omega_R = \omega_{SM}^2/\gamma < \omega_{SM}$ with the linear critical slowing down (10).

In many cases the dynamic anomalies around ferroelectric phase transitions are more complicated and contain features of both displacive and order-disorder behaviour. Typically, far above the transition, some phonon softening appears which ceases close to the transition point, where additional relaxational dispersion appears and usually contributes substantially to the permittivity maximum. Such a relaxation is frequently called a central mode in analogy with inelastic scattering experiments [9], where such excitations contribute as quasielastic peaks with a half-width equal to $\omega_R$. Such a situation is usually referred to as crossover from displacive to order-disorder behaviour.

In the ferroelectric phase, an additional contribution to permittivity as a rule stems from domain wall vibrations. It can usually also be modelled by a Debye relaxation (typically in the MHz-GHz range) which slows down on cooling according to the Arrhenius or Vogel-Fulcher law, resembling in this way a glassy or dipolar-glassy behaviour.

Let us now briefly comment on some specific features of high-permittivity polycrystalline materials. It is well established that even high-density cubic ceramics (~98% theor. density, e.g. for $BaTiO_3$ (BTO) [10] and $SrTiO_3$ (STO) [11, 12]), in which the porosity effect might be neglected, always show a reduced permittivity compared to single crystals; and the reduction increases with decreasing grain size. It has become clear recently that this is not due to an intrinsic dielectric size effect in the grains, but rather due to a thin low-permittivity layer (so called dead or passive layer) at grain boundaries, whose microscopic origin could be probably a deficit of oxygen. However, the microscopic origin of dead layers has not yet been carefully addressed. The ceramics behave like a core-shell composites, whose effective dielectric response up to the IR frequencies can be reasonably approximated using appropriate EMA models, like a brick-wall [11] or coated spheres model [12, 13]. Qualitatively it appears clear that the reduction of permittivity should be accompanied by an effective stiffening of the soft mode (increase in its frequency) which accounts for the reduced permittivity. We shall discuss the results obtained on particular materials in more detail below.

Concerning understanding of (even more pronounced) suppressing of the dielectric response in polycrystalline thin films, the situation appears more complicated because, in addition of the effect of grain boundaries, one has to consider also the interfacial layers between the film and substrate/electrode and the stresses exerted by the substrate on the film. A detailed analysis and separation of all these effects has, to our knowledge, not yet been performed for any high-permittivity film.

## 2. Ferroelectric, antiferroelectric and incipient ferroelectric materials

### 2.1 Perovskites

From the point of view of functionality, the best known and most applied ferroelectrics, antiferroelectrics and incipient ferroelectrics have perovskite structure ($ABO_3$ stoichiometry). Let us start with the best known incipient ferroelectric STO. In single crystals the permittivity increases from ~300 at 300 K to more than 24,000 at 5 K not reaching any ferroelectric transition (but undergoing - without any dielectric anomaly - an antiferrodistortive transition into a tetragonal phase near 105 K). However, in ceramics this increase is radically suppressed depending on the grain size, see Fig. 1. As seen from the fitted curves, all the data can be accounted for by a universal grain-boundary dead layer, but its thickness and local permittivity are strongly correlated and therefore cannot be determined independently. The measured IR reflectivity in these ceramics confirmed that the soft modes in ceramics are accordingly stiffened compared to the single crystal, see Fig. 2, and successfully explain the static permittivity in Fig. 1 [12]. The measured IR reflectivity of a nanograin (NG, grain size ~80 nm) ceramics was well fitted with a more complex doubly-coated spheres model assuming the unchanged single-crystal dielectric function for the bulk grain. The outer distorted grain-boundary dead layer is very narrow (~0.5 nm) having temperature independent dispersionless low permittivity ~10 and the inner thicker layer (~5 nm with only slightly distorted perovskite structure) accounts for the polar properties in the STO ceramics, deduced from the appearance of the forbidden IR modes in the Raman spectra [14].

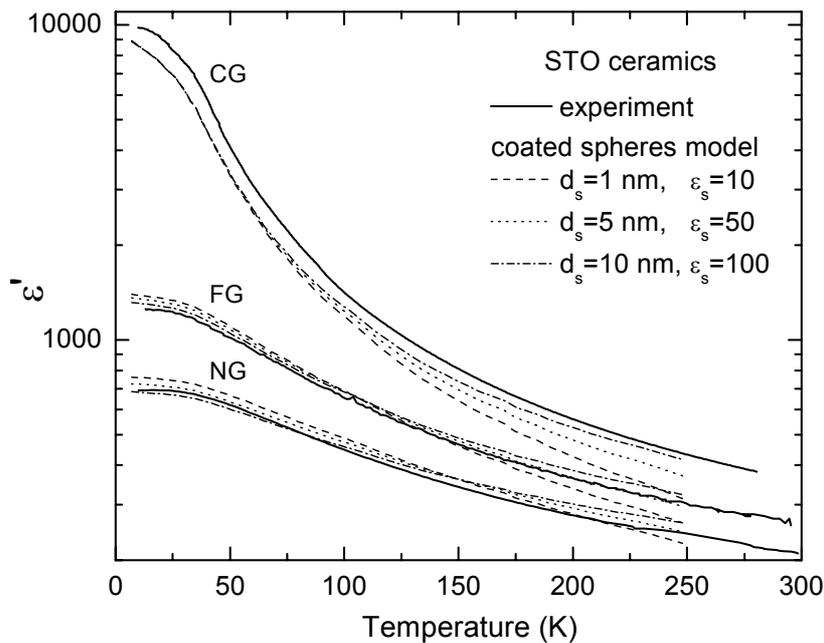

Fig. 1. Low frequency dielectric data of the studied STO ceramics. The coarse-grain ceramics CG - mean grain size 1500 nm, fine-grain ceramics FG - 150 nm, nanoceramics NG – 80 nm. No dispersion in any of the samples was seen for 100 Hz-1 MHz. The data are compared with the calculation from the coated spheres model using various dead layer widths $d_s$ and static permittivity $\varepsilon_s$. Due to the strong correlation of the two quantities, these parameters cannot be determined from the static permittivity data alone fulfilling $\varepsilon_s/d_s = 10$. (Reproduced from [12], p. 231).

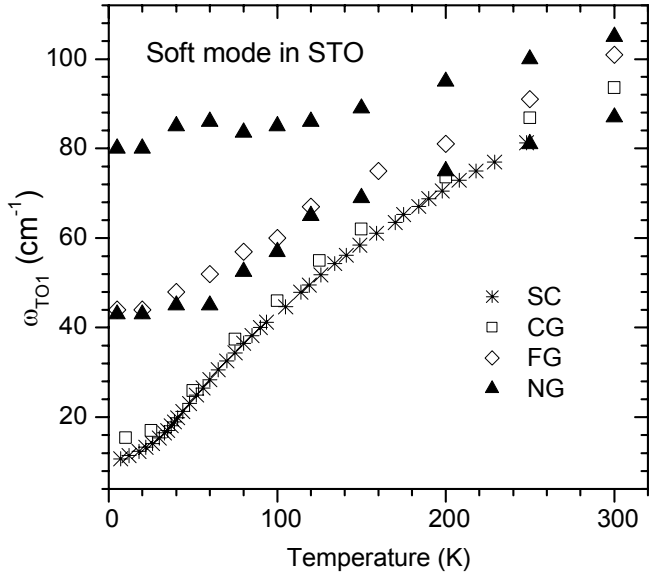

Fig. 2. Temperature dependence of the soft mode frequencies in the studied ceramics compared with that in single crystal SC. Only in the NG ceramics the soft mode response was fitted with 2 oscillators, whereas the single dielectric loss maximum lies in between them. (Reproduced from [12], p. 236).

Careful Raman studies on the same STO ceramics [11, 12, 14] yield the low-temperature ferroelectric soft-mode frequency of ~23 cm$^{-1}$; unlike the IR results, which yield ~44 and ~60 cm$^{-1}$ for the FG and NG ceramics, respectively. This can be explained by the different probing wavelengths used in both the techniques. In the case of IR technique the wavelength is much larger than the grain size so that the EMA approximation, which requires a homogeneous *E* field inside individual grains, is applicable. In the case of Raman technique, however, the two lengths are comparable and the EMA approximation is no longer justified [11]. From the physical point of view, in the latter case the effect of depolarizing field (fully accounted for by EMA) on the soft mode is not as strong as in the former case. It shows that the two techniques can yield slightly different results in the case of inhomogeneous samples, which has been so far not taken into account.

In the case of ferroelectric BTO, the grain-size effect on the dielectric response was studied at the earliest [15]; but in the ferroelectric phase it is determined mainly by the domain-wall contribution and therefore by the domain structure in the grains, which is strongly grain-size dependent. We shall limit ourselves on the discussion of the paraelectric phase, where the dielectric size effect almost cancels the Curie-Weiss dependence of the permittivity [10]. Recently, it was confirmed by the IR reflectivity measurements (down to 50 nm grain size) that also here the suppression of permittivity is due to the soft mode stiffening [16]. Raman data on the same set of nanoceramics [17] confirmed that all three ferroelectric phase transitions take place, at least locally, since the Raman technique is a rather local probe (correlation length of few nm, shorter than e.g. for XRD). On the other hand, it was shown [18] that a free BTO single crystal platelet with the thickness of 75 nm shows no reduction of permittivity and no smearing or shift of the dielectric anomaly near $T_C$ observed in BTO nanoceramics. This confirms the absence of any size effect in BTO down to 75 nm, in agreement with first-principles based calculations which show that the intrinsic size effect for the BTO film thickness starts in the range of a few lattice constants only [19]. Therefore, the dielectric size effect in (nano)ceramics is predominantly due to the grain-boundary dead layers, and smearing of the dielectric anomaly near $T_C$ in thin films can also be due to an inhomogeneous strain in the film by the substrate clamping of the film [18].

The classical antiferroelectric $PbZrO_3$ (PZO) shows a single antiferroelectric transition at 508 K, which is thermodynamically very close to a ferroelectric transition and is connected with a strong Curie-Weiss anomaly of the permittivity. The temperature dependent IR reflectivity complemented with MW waveguide data at 36 GHz and data near 10 $cm^{-1}$ (approx. 300 GHz) by backward-wave-oscillator spectroscopy (at low temperatures) on dense ceramics revealed near but above $T_C$ a strong heavily damped TO mode near 50 $cm^{-1}$ and, additionally, an overdamped mode in the 10 $cm^{-1}$ range (central mode), whose temperature dependence accounts for the main part of the dielectric anomaly – see Fig. 3 [20]. As seen in Fig. 3, below $T_C$ the central mode quickly vanishes and the whole permittivity is due to the phonon contribution. Somewhat later, another dense (98% theor. density) and hot-pressed single-phase PZO ceramic, was studied, which showed much smaller dielectric anomaly (maximum permittivity below 1000), but with the same $T_C$ and similar phonon spectrum [21]. The difference turned out to be due to nano-cracks along some of the grain boundaries. The brick-wall model calculation showed that 0.29 vol. % of the crack porosity is enough to account for the whole effect, which is due to hardening of the central mode by a factor of 4 in the 10 GHz range. Such a pronounced effect points to the need for some caution in interpreting the dielectric data of high-permittivity ceramics.

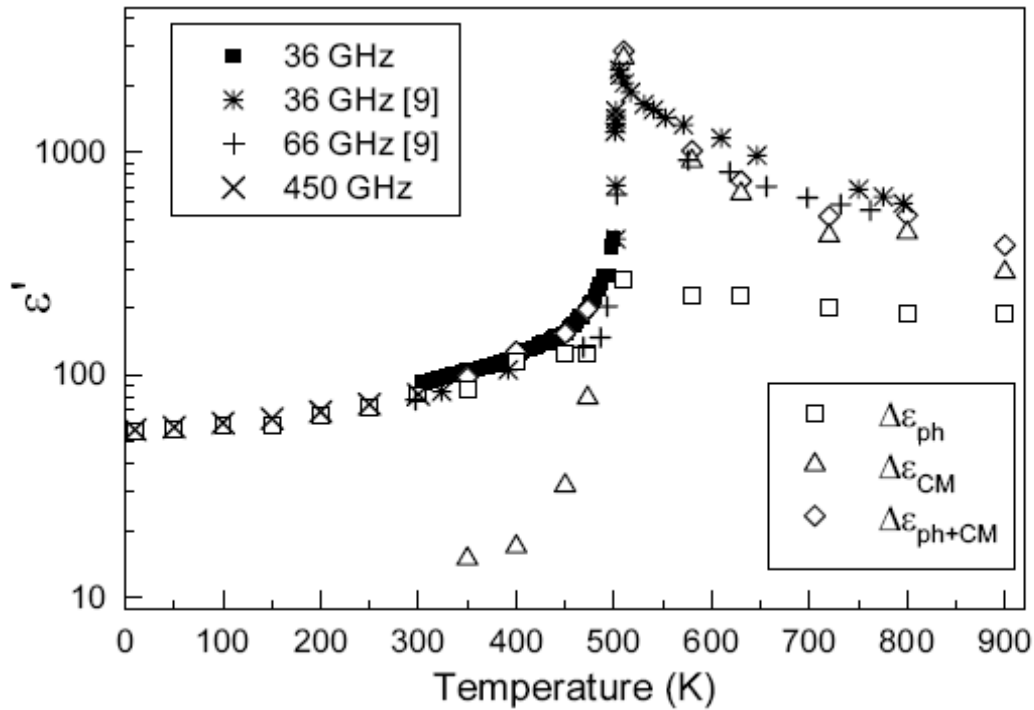

Fig. 3. Temperature dependent contributions to the permittivity due to the phonon modes ($\Delta\varepsilon_{ph}$), central mode ($\Delta\varepsilon_{CM}$) and the sum of both contributions ($\Delta\varepsilon_{ph+CM}$) which accounts for the whole low-frequency permittivity. (Reproduced from [20], p. 2685).

The doping of incipient ferroelectrics by various metals usually induces dipolar glassy behaviour which on increasing doping changes into ferroelectric phase transitions [22]. This is usually accompanied by low-temperature dielectric relaxations, but it also influences the soft phonon behaviour. This is expected, since the pure systems are close to the ferroelectric instability. Among them, the closest to it is STO (the highest low-temperature permittivity among incipient ferroelectrics), to which most of the work has been devoted. So far, the effects of Ca, Ba, Pb, Bi, La, Mg, Mn, and Nb dopants in STO have mainly been studied. Recently, two types of doping were attempted, substituting either Ti (perovskite B-sites) or Sr (A-sites) by Mg [23] and Mn [24] and different dielectric and soft mode behaviour was revealed. A-site doping seems to induce a ferroelectric transition and an additional low-frequency – low-temperature relaxation, whereas B-site doping drives the system away from ferroelectricity (smaller permittivity, stiffening of the soft mode). However, the confidence in the obtained results was questioned [25] and presently a study on similarly doped single crystals, which yield somewhat different results, is in progress.

The best studied solid solution based on STO is BTO-STO (BST), which exists in the whole concentration range and for Ba concentration above ~3.5%, it displays a rather sharp ferroelectric transition with $T_C$ varying with Ba concentration (linearly above ~10% Ba) from ~20 up to ~400 K [26]. The Raman spectroscopic results on BST (mainly on epitaxial thin films and single crystals) have been recently thoroughly reviewed [27]. The essential message is that the films differ from single crystals showing relaxor ferroelectric properties with polar nanoregions in the paraelectric phase, smearing of the ferroelectric phase transitions, suppression of the lower-temperature ferroelectric transitions and changes in the corresponding soft mode behaviour. The IR results on bulk BST ceramics [28, 29] show that the phonon contribution with a more complex soft mode behaviour accounts for the paraelectric permittivity except for the temperature range of ~20 K above $T_C$, where a dynamic central mode in the 10 GHz range is probably required to contribute to the sharp and pronounced low-frequency dielectric anomaly. With increasing Ba concentration the damping of the soft mode increases and its spectral line-shape near $T_C$ consists of two peaks in agreement with calculations based on the first-principles model Hamiltonian with one degree of freedom [29].

## 2.2. Aurivillius compounds

So far we have discussed only materials with perovskite structure, which have been up to now of the greatest interest from the point of view of applications. However, interest in ferroelectrics with high spontaneous polarization $P_S$ and $T_C$ and smaller fatigue (reduction of $P_S$ after long-term switching) turned the attention to high-temperature Bi-layered ferroelectrics with an Aurivillius structure, whose structure consists of perovskite-like blocks interleaved with $Bi_2O_2$ layers. The best studied compounds, including recent studies using THz spectroscopy, are $SrBi_2Ta_2O_9$ (SBT) (already utilized in FERAM memories) [30] and $Bi_4Ti_3O_{12}$ [31]. THz spectroscopy revealed in these compounds a well underdamped low-frequency polar mode near 30 cm$^{-1}$, which only slightly softens on heating, even across $T_C$, and does not explain the low-frequency dielectric anomaly. Another lower frequency relaxation (central mode), whose high-frequency tail in the case of SBT was directly seen in the THz spectra [30], is needed to explain the dielectric anomaly. Since no structural disorder was revealed in these compounds, the central mode dynamics was assigned to dynamic polar clusters in the paraelectric phase and the ferroelectric transition to a percolation transition of these clusters (contact of clusters along a macroscopic path), where some of them merge into macroscopic ferroelectric domains. Such a transition is neither purely displacive (as earlier assumed for these compounds) nor purely order-disorder in the local sense.

## 2.3 Pyrochlores

Another structural type of interest concerning the dielectric properties and ferroelectricity are pyrochlores of the $A_2B_2O_7$ basic stoichiometry [32]. They often appear as an unwanted second phase when processing perovskites. The best studied pyrochlore is $Cd_2Nb_2O_7$, which undergoes seven phase transitions below 512 K including a ferroelectric one at 196 K accompanied by a soft phonon mode observed in the IR as well as in Raman spectra and a 'central-mode type' dielectric dispersion in the near-mm range [33]. It was shown that its relaxor properties in the ferroelectric phase are apparently due to domain wall dynamics [33]. Another interesting pyrochlore, appearing frequently as an unwanted second phase in the famous relaxor $PbMg_{1/3}Nb_{2/3}O_3$ (PMN) ceramics or thin films, is $Pb_{1.83}Mg_{0.29}Nb_{1.71}O_{6.39}$. It was recently studied using IR, THz, MW and low-frequency dielectric spectroscopy [34] and was shown to exhibit an incipient-ferroelectric behaviour with a well underdamped soft mode in the 30 cm$^{-1}$ range (see Fig. 4) and permittivity increasing from ~90 at 1000 K to ~240 below 40 K. The soft mode frequency as well as the permittivity levels off below ~50 K and the soft-mode frequency follows the quantum paraelectric behaviour

$$\omega_{SM}(T) = \sqrt{A\left[\left(\frac{T_1}{2}\right)\coth\left(\frac{T_1}{2T}\right) - T_0\right]} \qquad (11)$$

with $T_1$ = 96 K and $T_0$ = -240 K, compatible with the Barrett formula for the permittivity [34, 35] assuming temperature independent soft mode oscillator strength.

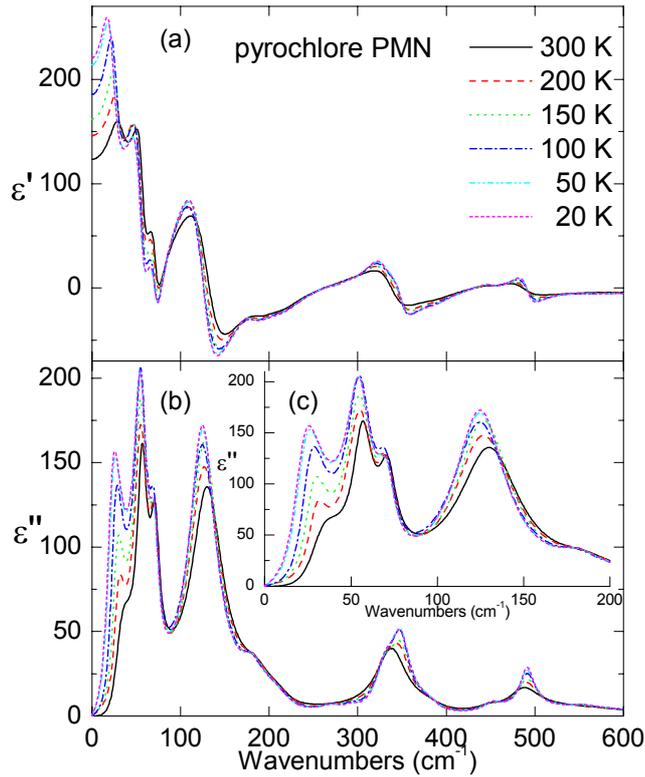

**Fig. 4**. (colour online) Dielectric function of the pyrochlore PMN ceramics from the fit to IR reflectivity and THz transmission data. Note the incipient ferroelectric behaviour. (Reproduced from [34], p. 054125-4).

## 3. Relaxor Ferroelectrics

### 3.1 Perovskites

Relaxor ferroelectrics (see Introduction) are nowadays intensively studied because they show very specific dielectric behaviour, interesting for applications; namely extremely high permittivity, electrostrictive and piezoelectric (under applied electric field) constants, and also for basic research concerning the complex and strongly temperature dependent dielectric dispersion, and other properties connected with the polar nanoregions (clusters). The dielectric dispersion was studied, for the first time in a sufficiently broad frequency and temperature region [36], on $(Pb_{1-x}La_x)(Zr_yTi_{1-y})O_3$ (PLZT100·x/y/1 − y) ceramics with x = 8 and 9.5 % and y = 65. Even if a low-frequency phonon mode was revealed which exhibited softening on heating, the phonon contribution to permittivity remained below ~500, and only at the highest temperature near Burns temperature $T_B$ = 620 K, where the polar regions are formed, did it contribute by the full value of the low-frequency permittivity. Below $T_B$ a MW relaxation splits off from the soft mode, on cooling it slows down and broadens considerably contributing by a substantial part to the permittivity. The central and low-frequency part of the dielectric-loss spectra slow down according to the Vogel-Fulcher law, explaining the characteristic frequency-dependent maximum of permittivity vs. temperature dependence. It was assigned to the polar cluster dynamics, at high temperatures probably mainly to flipping of their dipole moment and at lower temperatures mainly to breathing of the clusters (fluctuations of their volume due to cluster-wall dynamics) [36]. However, the high-frequency part (GHz-THz range) of the loss spectra below the freezing temperature (comparable to the Vogel-Fulcher temperature $T_{VF}$ near 230 K for PLZT) is much less temperature dependent and the losses form frequency-independent spectra from ~$10^{10}$ Hz down to the lowest measurable frequencies (so called 1/f or flicker noise) whose rather high level decreases exponentially with decreasing temperature. It was assigned to a very broad distribution of relaxation frequencies (much broader than the measured frequency range) and explained by a broad distribution (down to almost zero) of energy barriers for hopping of the off-centred Pb ions (main origin of the cluster dipole moment) near the cluster boundaries [37].

Qualitatively similar behaviour was also observed in all other investigated perovskite relaxors: B-site disordered PMN ($PbMg_{1/3}Nb_{2/3}O_3$) (i.e., Mg and Nb ions are rather randomly distributed over the $ABO_3$ perovskite B-sites, even if some local chemical ordering may exist) [38, 39], B-site ordered and disordered PMT ($PbMg_{1/3}Ta_{2/3}O_3$) [40], and B-site ordered and disordered PST ($PbSc_{1/2}Ta_{1/2}O_3$) [41]. The ordered forms show doubling of the unit cell due to regularly alternating B-ions. Even if the dielectric response is sensitive to the degree of B-site ordering and to the type of sample (single crystal, bulk ceramics, thin film), the differences are revealed only in the lower-frequency relaxational dynamics, but almost never in the phonon spectra. On cooling from high temperatures, the overdamped soft phonon mode slightly softens to $T_B$. Below that, it splits into a stiffer underdamped component polarized along the local cluster dipole moment, which hardens on cooling following the Cochran law (up to ~80 cm$^{-1}$ at low temperatures), and a much softer almost overdamped low-frequency component, which hardly changes with temperature and remains in the 30 cm$^{-1}$ region down to cryogenic temperatures. It is assigned to vibrations whose dipole moment is polarized perpendicular to the local dipole moment of polar clusters. In the case of PST, PMN-PT and PZN-PT, in which a first-order ferroelectric transition appears on cooling, no phonon softening is connected with this transition [41, 42]. A rather thorough analysis of the polar phonon behaviour in perovskite relaxors was recently performed by Hlinka et al. [4]. These features are not discussed in detail here.

One common feature of perovskite relaxors and ferroelectric perovskite ceramics with anisotropic grains is worth commenting upon. It has already been mentioned that the soft polar phonon is strongly split below $T_B$ into two components, polarized along and perpendicular to $P_s$, which results from a strong uniaxial anisotropy of the local dielectric function. The strong dielectric anisotropy is known from single-domain single crystal studies (e.g. for tetragonal BTO [43]). This feature, which in a similar way to the relaxors influences the effective spectra of polycrystalline samples (ceramics, films), can be treated in a similar way, as long as the probing electric field in individual grains or clusters is homogeneous, i.e. the IR wavelength is much longer than the grain (cluster) diameter. The simplest model which can be used in this case is the Bruggeman EMA approximation, which was successfully used for fitting the IR reflectivity of a PMN crystal to reveal the high local dielectric anisotropy [44]. It was used also for fitting several relaxor and normal ferroelectric ceramics [4, 45] and, earlier, for fitting polycrystalline PbTiO$_3$ films [46].

Recently a search for environmentally friendly relaxor materials, not containing lead, has become quite popular. One of the best investigated, but with a rather complex phase transition behaviour, is Na$_{1/2}$Bi$_{1/2}$TiO$_3$ [47], the only known relaxor with an A-site disorder (Na and Bi ions almost randomly distributed over the A-sites). Like in other relaxors discussed, it shows an overdamped low-frequency polar mode in the THz range, which partially softens on heating towards the temperature of permittivity maximum near 600 K (no higher temperature data are available). However, this temperature lies in between of two structural transitions on cooling from high temperatures: near 810 K (cubic-tetragonal) and 470 K (tetragonal-rhombohedral) with a broad region of coexistence of both phases in between. Both transitions are connected with a unit cell multiplication. Recently it has been shown [48] that the maximum in permittivity is connected with an appearance of incommensurate rhombohedral modulation along the tetragonal axis within the tetragonal phase, which then remains stable (with a periodicity of ~3.2 nm) down to low temperatures. This modulation as well as the dielectric anomaly is connected with a local chemical A-site order existing from the cubic phase. Again, phonons contribute only by 150-300 to the permittivity, the main contribution stemming from a central mode in the 10 GHz range [47]. Both these excitations are assigned to the strongly anharmonic vibrations of the off-centred A-cations.

**3.2 Tungsten bronzes**

The best studied relaxor system with tungsten-bronze structure is barium strontium niobate (Sr$_x$Ba$_{1-x}$)Nb$_2$O$_6$ (SBN). Since the system is tetragonal with a strong dielectric anisotropy and single crystals are available, it was studied mostly on single crystal samples. The dielectric response along both principal directions at different frequencies including the phonon contribution is shown in Figure 5 [49]. As is characteristic for relaxors, the $\varepsilon_c'(\omega)$ response in the paraelectric phase above $T_C \sim 360$ K shows a complex frequency dispersion (three relaxation regions below the phonon response) and the phonons do not contribute at all to the Curie-Weiss anomaly. Interestingly, the phonon contribution to $\varepsilon_a'$ is much larger than to $\varepsilon_c'$ even if the low-frequency permittivity anisotropy is the opposite; the polar cluster dynamics contributes much more to $\varepsilon_c'$.

Recently, some other tungsten-bronze ceramics with relaxor behaviour were synthesized and dielectrically characterized: $Ba_2(La,Nd)Ti_2Nb_3O_{15}$ (BLNTN) solid solution system [50] and $Sr_2LaTi_2Nb_3O_{15}$ (SLTN) [51]. The first system passes from purely relaxor behaviour without any phase transition for the La compound to a sharp first-order ferroelectric transition near 390 K with only small paraelectric dispersion up to 300 MHz for the Nd compound. The $La_{1/2}Nd_{1/2}$ compound shows a combination of both features. SLTN behaves in a purely relaxor-like way with no phonon softening and with absence of any ferroelectric phase transition even in an electric field. This is characteristic of dipolar glasses, rather than of relaxors. 20 polar modes were observed with very small temperature dependence as compared to 33 modes predicted by factor-group analysis for the paraelectric phase. The permittivities in BLNTN and SLTN ceramics are much smaller than that of SBN, and stay in the range of several hundreds.

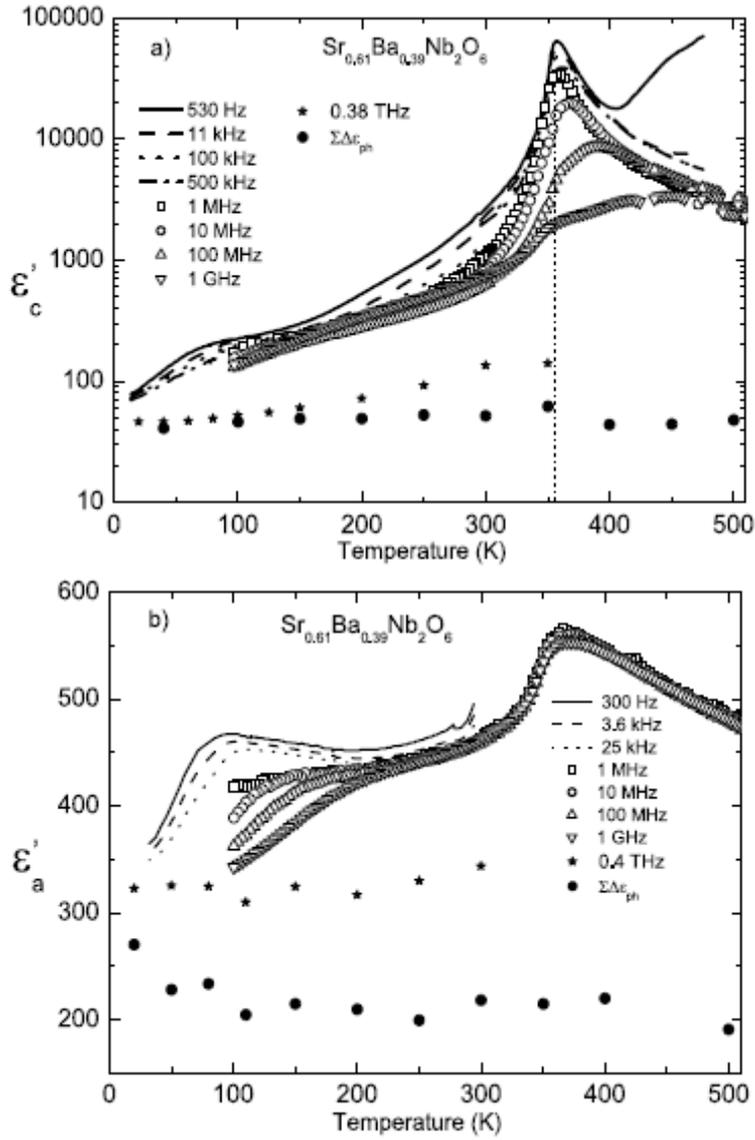

**Fig. 5.** Dielectric permittivity of SBN along and perpendicular to the tetragonal axis *c* at different frequencies. Note the logarithmic scale of $\varepsilon_c$'. (Reproduced from [49], p. 662).

## 4. Magnetoelectric multiferroics.

### 4.1. BiFeO$_3$

Study of magnetoelectric multiferroics,( i.e. the materials which exhibit simultaneously coupled ferroelectric and magnetic order), has become a hot topic in recent years. This is not only due to fascinating physics of magnetoelectric coupling, but also due to high potential of magnetoelectrics for technical applications. Such materials are highly promising for new generation of various sensors, actuators, and MW phase shifters. But the most challenging application is for their use in random access memories (RAMs). In such systems the information can be written by electric field and read non-destructively by magnetic sensing. These memories avoid the weak points of the ferroelectric RAMs (destructive reading causes fatigue) as well as of magnetic RAMs (high electric current is needed for overwriting, which rules out high integration of magnetic RAMs). Recently, it was actually demonstrated that a thin 2 nm magnetoelectric $La_{0.1}Bi_{0.9}MnO_3$ film can be used as a tunable spin filter in RAM's at 80 K.[52, 53] Unfortunately, there are not many magnetoelectric single-phase multiferroics known up to now and only a few of them have both magnetic and ferroelectric critical temperatures above room temperature.[54, 55] Therefore, there is an intensive search for new magnetoelectric multiferroic materials with high magnetization and spontaneous polarization above room temperature. For their discovery a detailed understanding of the mechanism of magnetoelectric coupling is necessary.

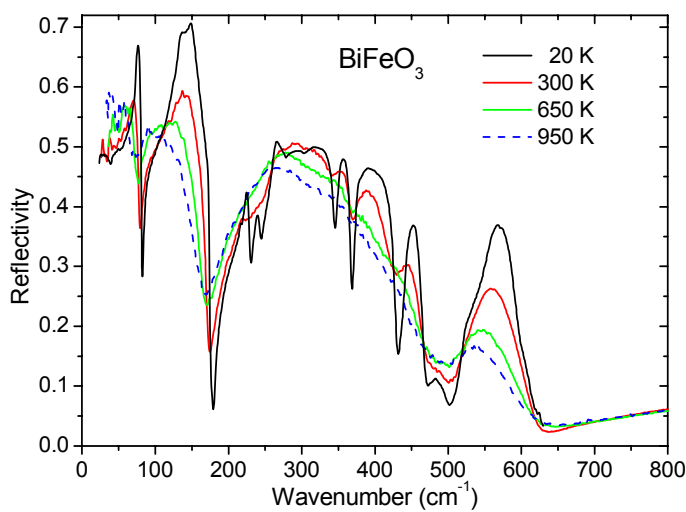

**Fig. 6.** (colour online) IR reflectivity spectra of $BiFeO_3$ ceramics at various temperatures. 13 polar modes observed at low temperatures correspond to factor group analysis in the rhombohedral R3c structure. 3 IR active modes are allowed in the cubic $Pm\bar{3}m$ structure above 1100 K, therefore many reflection bands gradually disappear on heating and finally only 4 broad reflection bands at 950 K are seen. (Reproduced from Ref. [57], p. 024403).

A classical high-temperature magnetoelectric multiferroic, found already at the beginning of 1960's, is BiFeO$_3$ (BFO) [56]. It exhibits an antiferromagnetic phase transition at $T_N$ near 640 K and a ferroelectric phase transition at $T_C \approx 1120$ K. Unfortunately, BFO frequently exhibits defect conductivity (mainly due to oxygen nonstoichiometry) which rules out the capacitance measurements above room temperature. Giant extrinsic permittivity appears at higher temperatures due to nonzero conductivity in the bulk grain combined with higher resistivity of the grain boundaries. The same giant effective permittivity can be observed also in crystals, probably due to depletion layers near electrodes [6]. The combination of this so-called Maxwell-Wagner polarization mechanism and inhomogeneous magnetoresistance is responsible for the change of permittivity with magnetic field observed in BFO ceramics [57]. However, at high frequencies these spurious effects disappear. The IR and THz spectra of BFO ceramics up to 950 K [57] are not influenced by the conductivity. THz permittivity increases on heating due to a partial phonon softening observed in the IR spectra. The increase is small (permittivity value is less than 50 at 900 K), probably due to improper ferroelectric type of the phase transition [57], since the unit cell seem to double below $T_C$. Lobo et al. [58] investigated the IR spectra of a BFO single crystal and qualitatively confirmed the results obtained on ceramics, but with the double the permittivity contributed by phonons, due to predominant contribution of the E symmetry modes to the IR reflectivity. Raman spectra of BFO single crystals [59, 60] show gradual decrease of intensities of Raman modes on heating and their disappearance above 1100 K. This corresponds to expected change of the crystal symmetry to the cubic Pm$\overline{3}$m phase, but it can be also an effect of the phase transition to a conducting phase recently observed by Palai et al. [61]. Frequencies of phonons observed in the Raman and IR spectra correspond well with the theoretical values calculated from the first principles [62].

The incommensurate magnetic order in BFO induces a magnon (spin wave) branch folding in the reciprocal lattice space, which activates several Raman spin-wave modes with different wave vectors allowing investigation of spin waves by optical probes [63]. Magnons in BFO are by symmetry arguments linearly coupled to polar phonons, also contributing to the dielectric function, and are called electromagnons. Cazayous et al. [64] revealed in their Raman experiment two-magnon branches, corresponding to spin wave excitations in and out of the cycloidal magnetic plane.

**4.2. $BiFe_{1/2}Cr_{1/2}O_3$**

Baetting and Spaldin [65] predicted from *ab initio* calculations that the chemically ordered double-perovskite $BiFe_{1/2}Cr_{1/2}O_3$ (BFCO) should have improved magnetic and ferroelectric properties compared with BFO. Suchomel et al. [66] prepared a BFCO ceramics, but the magnetic phase transition was revealed only below 130 K. However, Nechache et al. [67, 68] processed an epitaxial thin films and revealed both ferroelectric and antiferromagnetic properties at room temperatures. The discrepancy with the magnetic properties of ceramics was explained by the absence of ordering of Fe and Cr cations in ceramics. High-temperature magnetic measurements of BFCO thin films confirmed the magnetic phase transition above 600 K [69]. IR reflectivity spectra of the thin film revealed no structural phase transition up to 900 K, the sample staying ferroelectric at least up to this temperature. The phonon contribution to static permittivity increased on heating due to a softening of several phonons as for BFO [69]. It appears that the B-site ordered BFCO belongs to the rare case of high-temperature magnetoelectric multiferroics.

**4.3. EuTiO$_3$**

EuTiO$_3$ represents another interesting magnetoelectric material. It is not ferroelectric, but quantum paraelectric like SrTiO$_3$ or pyrochlore PMN. Its permittivity increases on cooling and saturates below 30 K [70]. EuTiO$_3$ exhibits an antiferromagnetic phase transition at $T_N$ = 5.5 K and below this temperature it's permittivity dramatically drops down due to coupling of the magnetic spins with crystal lattice [70]. In the antiferromagnetic phase, the permittivity increases by 7% with magnetic field giving evidence for a huge magnetoelectric coupling [70]. IR reflectivity and THz transmission spectra of ceramics revealed a polar optic phonon whose softening is fully responsible for the observed dielectric behaviour [71]. The soft mode frequency $\omega_{SM}$ follows the formula (eqn 11) which takes into account the zero-energy quantum fluctuations at low temperatures.

Since the permittivity can be tuned by a magnetic field, it is expected that the soft mode frequency should be dependent on the magnetic field. The first attempt to reveal at 1.8 K the soft phonon dependence on the magnetic field up to 13 T remained below the accuracy of the IR reflectivity experiments on ceramics [71]. Nevertheless, the same experiment performed on a 1% compressively strained EuTiO$_3$ thin film deposited on (LaAlO$_3$)$_{0.29}$-(Sr$_{1/2}$Al$_{1/2}$TaO$_3$)$_{0.71}$ revealed a small (2.5 cm$^{-1}$) but reliable magnetic tuning of the soft mode frequency [72].

**5. Microwave ceramics**

Unlike materials discussed so far, MW materials are, as a rule, non-ferroelectric materials with an only weakly anharmonic lattice, not undergoing structural phase transitions, and having only moderate permittivity of 20-100. They are mostly used as MW resonators. The permittivity should be almost temperature independent to assure as small as possible temperature variation of the dielectric resonance frequency and the dielectric losses should be also as small as possible to yield high-quality resonances [73]. The temperature independence of the permittivity is given by a small temperature dependence of the polar phonon frequencies and its frequency dependence in the MW range can be neglected at frequencies an order of magnitude below the phonon mode range (see eqn (2)). If the MW losses are intrinsic, they are given by the multiphonon (mostly two-phonon) absorption tails below the lowest TO mode [74]. To a first approximation they can be described by eqn (2), but this rather phenomenological model should be applied with caution [75], since the mode damping parameters are not necessarily frequency independent down to the MW range. Nevertheless the microscopic phonon kinetic theory [74] gives (at moderate temperatures) the same proportionality of losses to frequency $\varepsilon'' \propto \omega$ as eqn (2) sufficiently below the phonon mode frequencies. This can be used to estimate the intrinsic MW losses by a simple extrapolation from the losses in the THz range and for well-processed materials it agrees with the directly measured MW losses. Since the much higher THz losses are usually not as sensitive to technology as the lower MW losses, THz spectroscopy is a convenient tool to check the expected magnitude of MW losses and by comparing it with the measured MW losses to check on the optimization of the processing [76]. In [76] also a large table is given, which compares the MW and THz permittivities and quality factors $Q \times \omega$ ($Q = \varepsilon'/\varepsilon''$) for about 60 most frequently used and studied ceramics. The agreement between THz and MW data is in most cases quite satisfactory.

In Fig. 7 we give an example of the broad-range dielectric spectra of a novel complex-perovskite ceramic materials $Ba(Zn_{1/3}Nb_{2/3})O_3$-$Ba(Ga_{1/2}Ta_{1/2})O_3$ (BZN-BGT) [77]. It is seen that, whereas the permittivity shows no measurable dispersion below the rather complex phonon response, and the extrapolation agrees perfectly with the MW data, the losses (on a log scale) show some extra THz contribution in some of the samples, probably due to some charged defects.

Whereas at room temperature the losses seem to be in many cases almost intrinsic, the situation changes completely on cooling. Intrinsic losses should decrease on cooling quite steeply [74, 75], whereas the experiments show only a moderate decrease [76, 78]. The question of the origin of these extrinsic MW losses was, however, not yet carefully addressed for any material. It is a rather formidable task requiring broad MW-THz frequency loss data over a broad temperature range, and in close connection with variations of material processing.

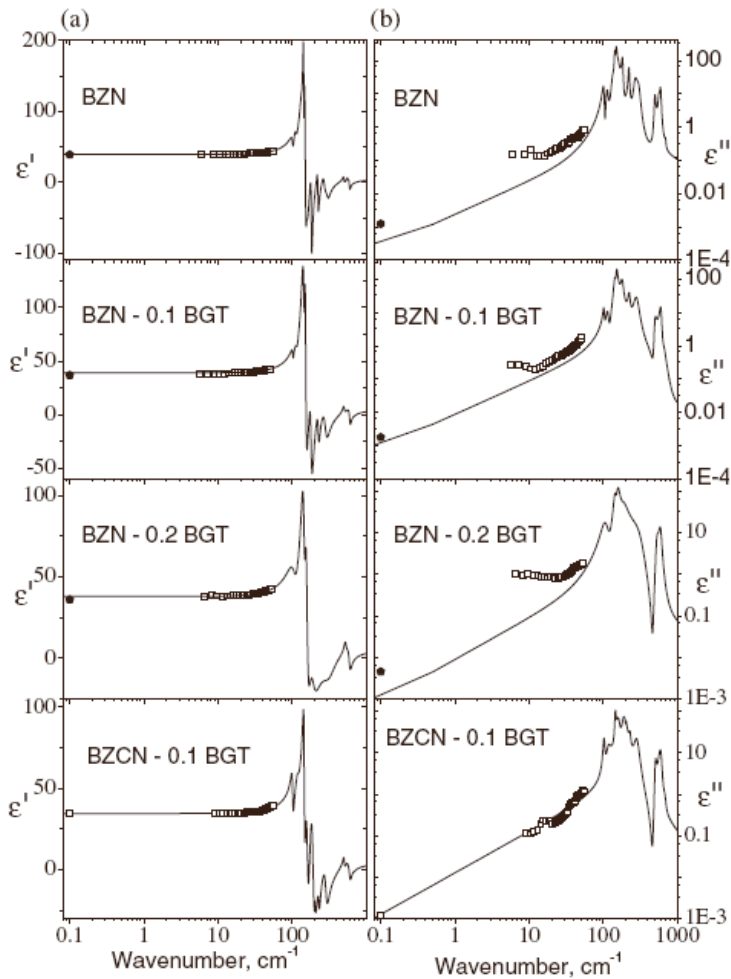

**Fig. 7.** Room temperature dielectric spectra of BZN-BGT ceramics. Comparison of IR, THz and MW data together with fit of IR reflectivity and its extrapolation to MW (full lines). Note the log scale for losses. ((Reproduced from [77], p. 1984).

**Acknowledgments**

The work was supported by the Czech Science Foundation (Project No. 202/06/0403) and by Academy of Sciences of the Czech Republic (AVOZ10100520).

**Glossary:**

**Phonon** – quantum of lattice vibrations in the crystal lattice;

**Polar phonon mode** – mode of lattice vibrations which displays a spatially (almost) uniform dipole moment enabling its coupling with an IR wave and so the first-order absorption process;

**TO mode** – transverse optic vibrational mode, in which the atoms vibrate perpendicularly to the propagation direction (wave vector) of the phonon wave; only TO modes can be polar;

**LO mode** – longitudinal optic vibrational mode, in which the atoms vibrate along the propagation direction of the phonon wave;

**Ferroelectric soft mode** – low-frequency polar phonon mode whose frequency tends to zero (softens) for temperature approaching the ferroelectric phase transition;

**Hard mode** – any mode which does not soften;

**Central mode** – excitation which usually appears in the microwave range close to the phase transition in addition to soft mode, as a rule of strongly anharmonic (not phononic) origin;

**Overdamped oscillator** – oscillator whose damping is so high that after deflecting it restores to equilibrium in a monotonic way;

**Underdamped oscillator** - oscillator whose damping is lower, i.e. it restores to equilibrium in damped oscillating way;

**Displacive phase transition** – structural phase transition in weakly anharmonic crystal lattice, where each atom occupies only a single site in the unit cell which slightly shifts due to the phase transition; dynamically it is characterized by appearance of a soft phonon mode;

**Order-disorder phase transition** - structural phase transition in a strongly anharmonic crystal lattice, where at least one type of atoms in the unit cell occupies at least two sites performing hops among them dynamically described by a critical Debye relaxation; the phase transition causes gradual ordering of these atoms in one of the sites;

**Antiferroelectric phase transition** – phase transition connected with an appearance of frozen antiparallel dipole moment in the unit cell, which by an applied electric field can be switched into a parallel arrangement inducing a ferroelectricity;

**Antiferrodistortive phase transition** – phase transitions connected with a doubling (or higher multiplication) of the unit cell; the corresponding critical excitation (soft mode) in the parent (high-temperature) phase is a phonon at Brillouin-zone boundary, which is therefore not optically active; it might become weakly active only in the distorted (low-temperature) phase due to folding of the Brillouin zone of the distorted phase;

**Incipient ferroelectric** - material which tends to ferroelectric transition under cooling, i.e. its permittivity increases (usually at not very low temperatures according to the Curie-Weiss law), but never reaches the transition point;

**Quantum paraelectric** – incipient ferroelectric in which the permittivity saturates and levels off at low temperatures due to quantum zero-point vibrations of the soft mode;

**Relaxor ferroelectric** – material which stays macroscopically nonpolar, but shows a pronounced and frequency dependent maximum in the temperature dependence of the permittivity, whose spectrum (frequency dependence) undergoes a complex relaxational dispersion;

**Improper ferroelectric** – ferroelectric in which the polarization is not the order parameter in the sense of Landau theory; at improper ferroelectric transition the (small) spontaneous polarization appears only as a secondary effect due to a coupling with the primary order parameter; the transition is connected only with a small dielectric anomaly and no Curie-Weiss law is expected to hold;

**Magnon** – quantum of spin waves in magnetically ordered crystals;

**Electromagnon** − magnon mode which is linearly coupled with the electric field of the IR wave and contributes (weakly) to the dielectric function;

**References**

[a] *Address: Institute of Physics, Academy of Sciences of the Czech Rep., Na Slovance 2, 182 21 Praha 8, Czech Republic, Fax: +420 286890415; Tel: +420 266052166; E-mail: petzelt@fzu.cz , kamba@fzu.cz*